\documentclass[twocolumn,prb,showpacs,superscriptaddress]{revtex4}
\usepackage{amsmath}
\usepackage{graphicx}
\usepackage{amssymb}
\usepackage{color}

\begin{document}

\title{Decay of Persistent Spin Helix due to the Spin Relaxation at Boundaries}

\author{Valeriy A. Slipko}
\affiliation{Department of Physics and Astronomy and USC
Nanocenter, University of South Carolina, Columbia, SC 29208, USA}
\affiliation{ Department of Physics and Technology, V. N. Karazin
Kharkov National University, Kharkov 61077, Ukraine }

\author{Anastasia A. Hayeva}
\affiliation{ Department of Physics and Technology, V. N. Karazin
Kharkov National University, Kharkov 61077, Ukraine }

\author{Yuriy V. Pershin}
\email{pershin@physics.sc.edu}
\affiliation{Department of Physics and Astronomy and USC
Nanocenter, University of South Carolina, Columbia, SC 29208, USA}

\begin{abstract}
We study electron spin relaxation in one-dimensional structures of finite length in the presence of Bychkov-Rashba spin-orbit coupling and boundary spin relaxation. Using a spin kinetic equation approach, we formulate boundary conditions for the case of a partial spin polarization loss at the boundaries. These boundary conditions are used to derive corresponding boundary conditions for spin drift-diffusion equation. The later is solved analytically for the case of relaxation of a homogeneous spin polarization in 1D finite length structures. It is found that the spin relaxation consists of three stages (in some cases, two) -- an initial D'yakonov-Perel' relaxation is followed by spin helix formation and its subsequent decay. Analytical expressions for the decay time are found. We support our analytical results by results of Monte Carlo simulations.
\end{abstract}

 \maketitle


Dynamics of electron spin polarization in semiconductor structures has attracted a lot of attention recently in the context of spintronics~\cite{Zutic04a,Bandyopadhyay08a,Wu10a},
 which is playing a fundamental role in the novel technological developments based on different effects in this scientific area. Moreover, the ability to understand and predict the dynamics of electron spins in semiconductors
is also important for the area of two terminal electronic devices with memory, so-called memristive devices~\cite{pershin08a,diventra09a,pershin11d,chua71a,chua76a}. In some of them~\cite{pershin08a,pershin11d}, the electron spin degree of freedom defines their internal state and, consequently, is responsible for their time-dependent memory response.

It has been shown by us recently~\cite{Slipko11a,Slipko11d,Slipko11c} that the system boundaries significantly modify the
dynamics of process. For example, we have demonstrated~\cite{Slipko11d} that in finite size 2D systems, the spin polarization density decays much slower than in the bulk and the exponential spin polarization decay rate is
defined by both the system size and strength of spin-orbit interaction. In finite length wires~\cite{Slipko11a} and
channels~\cite{Slipko11c} (oriented in a specific direction), changes in the electron spin relaxation are even more pronounced:
instead of relaxing to zero, the homogeneous electron spin polarization relaxes into a persistent spin polarization structure known as the
spin helix~\cite{Bernevig06a,Weber07a,Koralek09a,pershin10a} -- a spin polarization configuration in which the direction of spin polarization density rotates along the wire.

In real experimental situations, the spin helix configuration can not exist infinitely long. Here, we assume that the main decay mechanism is due to
spin relaxation at system boundaries. Indeed, local strong random electric fields in the vicinity of boundaries result in a random spin-orbit interaction influencing the electron spin degree of freedom. It is thus important to develop a theory and model spin relaxation in constrained geometries taking into account the boundary spin relaxation and understand how the boundary spin relaxation changes the overall character of electron spin relaxation in the entire system.

In this paper, we use both spin kinetic~\cite{Slipko11c} and diffusion~\cite{Burkov04a,Mishchenko04a,Saikin04a,Pershin04b,pershin10a} equations to investigate the dynamics of electron spin polarization in semiconductor wires of finite length. Specifically, we consider dynamics of spin relaxation in one-dimensional (1D) finite length systems with Bychkov-Rashba~\cite{Bychkov84a} spin-orbit interaction and boundary spin relaxation. Since it is easier to incorporate the boundary spin relaxation into the boundary conditions for spin kinetic equations~\cite{Slipko11c}, below, we use the spin kinetic equation approach first. Next, based on boundary conditions for the spin kinetic equations, we derive boundary conditions for spin diffusion equation which is easier to solve.  Finally, we obtain an exact solution for the problem of spin relaxation in finite length wire. Our analytical studies are complemented by semiclassical Monte Carlo simulations of spin dynamics~\cite{Saikin05a} giving an additional insight into the problem.


A kinetic description of electron spin polarization in one-dimensional wires can be given~\cite{Slipko11c} in terms of vectors
$\mathbf{S}^+$ and $\mathbf{S}^-$, which are the spin polarizations of electrons  moving along the wire in the positive (with momentum $\mathbf{p}=mv\mathbf{e}_x$), and negative ($\mathbf{p}=-mv\mathbf{e}_x$) $x$-directions with the average velocity $v=l/\tau$, where $l$ is the mean free path and $\tau$ is the momentum relaxation time. The kinetic equation can be written as a system of two vector equations~\cite{Slipko11c}
\begin{eqnarray}
\left(\frac{\partial }{\partial t}+v\frac{\partial }{\partial x}\right)
\mathbf{S}^+=-\Omega\mathbf{e}_y\times\mathbf{S}^+ -\frac{1}{2\tau}(\mathbf{S}^+-\mathbf{S}^-),
\label{KinEqPlus}  \\
\left(\frac{\partial }{\partial t}-v\frac{\partial }{\partial x}\right)
\mathbf{S}^-=\Omega\mathbf{e}_y\times\mathbf{S}^- -\frac{1}{2\tau}(\mathbf{S}^--\mathbf{S}^+),
\label{KinEqMinus}
\end{eqnarray}
which take into account electron spin precessions (first term in the r.h.s. of Eqs. (\ref{KinEqPlus}), (\ref{KinEqMinus})) induced by
Bychkov-Rashba~\cite{Bychkov84a} spin-orbit interaction and bulk scattering events (second term in the r.h.s. of Eqs. (\ref{KinEqPlus}), (\ref{KinEqMinus})).
Here, $\Omega=2\alpha p/\hbar$ is the angular velocity (directed along $y$-axis) of spin precessions, $\alpha$ is the
spin-orbit coupling constant, $p=mv$ is the average momentum, $\mathbf{e}_y$ is the unit vector in $y$-direction. Additionally, it is convenient to introduce
the parameter $\eta=\Omega/v=2\alpha m \hbar^{-1}$, which gives the spin precession angle per unit
length. Note that the direction of spin precessions in Eqs. (\ref{KinEqPlus}), (\ref{KinEqMinus}) is different for left- and right-moving electrons.

Eqs. (\ref{KinEqPlus}), (\ref{KinEqMinus}) incorporate the well-known D'yakonov-Perel' spin relaxation
mechanism~\cite{Dyakonov72a,Dyakonov86a}. Indeed, trajectories of spin-polarized electrons are randomized bulk scattering events
described by second term in the r.h.s. of Eqs. (\ref{KinEqPlus}), (\ref{KinEqMinus}).
Correspondingly, the direction of spin rotation becomes
fluctuating what causes average spin relaxation (dephasing). The incomplete spin relaxation in finite-length
wires~\cite{Slipko11a} can be explained by the existence of a maximum electron spin precession angle defined by the system size.

In this work, we study the dynamics of spin polarization in a wire of the length $L$, $-L/2<x<L/2$, in the presence of spin relaxation at the boundaries
$\Gamma=[x=\pm L/2]$. The boundary conditions, which take into account the relaxation of
spin polarization in elastic scatterings of electrons from the boundaries, can be formulated as
\begin{eqnarray}
\left[\mathbf{S}^-=\gamma
\mathbf{S}^+\right]|_{x=L/2},
\label{BCmp}\\
\left[\mathbf{S}^+=\gamma
\mathbf{S}^-\right]|_{x=-L/2},
\label{BCpm}
\end{eqnarray}
where $\gamma$, $0\leq\gamma\leq 1$, is a phenomenological dimensionless
papameter characterising the boundary relaxation rate.  If $\gamma=1$, then the spin polarization of scattered electrons is conserved (no boundary spin relaxation).  The case $\gamma=0$  corresponds to the total relaxation of spin polarization at the boundary, whereas for $0<\gamma< 1$ we deal with a
situation of a partial boundary relaxation.

Combining Eqs. (\ref{KinEqPlus}), (\ref{KinEqMinus}) we obtain a single equation
for the total spin polarization $\mathbf{S}=\mathbf{S}^++\mathbf{S}^-$
\begin{eqnarray}
\frac{\partial^2 \mathbf{S}}{\partial t^2}+\frac{1}{\tau}\frac{\partial \mathbf{S}}{\partial t}
-v^2\frac{\partial^2 \mathbf{S}}{\partial x^2}&-&2\Omega v \mathbf{e}_y\times\frac{\partial\mathbf{S}}{\partial x} \nonumber \\
&+&\Omega^2\left(\mathbf{S}-S_y\mathbf{e}_y \right)=0,
\label{KinEqSFin}
\end{eqnarray}
where $S_y$ is the $y$-component of $\mathbf{S}$. What are the boundary conditions for $\mathbf{S}$?
These can be derived reformulating the boundary conditions given by Eqs. (\ref{BCmp}),
 (\ref{BCpm}) in terms of the function $\mathbf{S}$ only. Subtracting Eq.
(\ref{KinEqMinus})  from Eq. (\ref{KinEqPlus})
we find
\begin{eqnarray}
\frac{\partial}{\partial t}\left(\mathbf{S}^{+}-\mathbf{S}^{-}\right)+ v\frac{\partial \mathbf{S}}{\partial x}
+\Omega\mathbf{e}_y\times\mathbf{S}+\frac{1}{\tau}\left(\mathbf{S}^{+}-\mathbf{S}^{-}\right)=0.
\label{KinEqDelta}
\end{eqnarray}
Now, let us consider, for example, the boundary $x=L/2$. From Eq. (\ref{BCmp}) we see that
at $x=L/2$, $\mathbf{S}^{+}=\mathbf{S}/(1+\gamma)$ and $\mathbf{S}^{-}=\gamma\mathbf{S}/(1+\gamma)$.
 Substituting these boundary values into Eq. (\ref{KinEqDelta}) we obtain
the corresponding boundary condition that can be generally formulated for both boundaries as
\begin{eqnarray}
\left.\left(v\frac{\partial \mathbf{S}}{\partial x}
+\Omega\mathbf{e}_y\times\mathbf{S}\pm\frac{1-\gamma}{1+\gamma}\left(\frac{\partial \mathbf{S}}{\partial t}+\frac{ \mathbf{S}}{\tau}\right)\right)\right|_{x=\pm
L/2}=0.
\label{BCS}
\end{eqnarray}

It is important to keep in mind that $S_y$ is not coupled to any other component of spin polarization (see Eqs. (\ref{KinEqSFin}), (\ref{BCS})).
Consequently, we can safely take out $S_y$ from our consideration selecting $S_y(x,t=0)=0$.

Introducing a complex polarization
\begin{eqnarray}
S=S_x+iS_z,
\label{ComplexS}
\end{eqnarray}
it is straightforward to rewrite  Eq. (\ref{KinEqSFin}) and boundary conditions (\ref{BCS})
in a more compact form
\begin{eqnarray}
\frac{\partial^2 S}{\partial t^2}+\frac{1}{\tau}\frac{\partial S}{\partial t}
-v^2\left(\frac{\partial}{\partial x}-i\eta\right)^2 S=0,
\label{ComplexSEq} \\
\left. \left(\frac{\partial S}{\partial x}-i\eta S\pm\frac{1-\gamma}{1+\gamma}\left(\frac{1}{v}\frac{\partial S}{\partial t}+\frac{ S}{v\tau}\right)\right)\right|_{x=\pm
L/2}=0.
 \label{ComplexSBC}
\end{eqnarray}

The diffusion limit of Eqs. (\ref{ComplexSEq}) and (\ref{ComplexSBC}) is realized
when $L\gg l$, $\eta l\ll 1$, and $t\gg\tau$. In  this case we can neglect $\partial^2 S/ \partial t^2$ compared to $\tau^{-1}\partial S/ \partial t$ in Eq. (\ref{ComplexSEq}) and
$\partial S/ \partial t$ compared to $S/ \tau $ in Eq. (\ref{ComplexSBC}),
because all relevant spin relaxation times are
much longer than  the momentum relaxation time $\tau$. As a result, we obtain
\begin{eqnarray}
\frac{\partial S}{\partial t}
=\frac{l^2}{\tau}\left(\frac{\partial}{\partial x}-i\eta\right)^2 S,
\label{ComplexSDEq} \\
\left. \left(\frac{\partial S}{\partial x}-i\eta S\pm\frac{1-\gamma}{1+\gamma}\frac{ S}{l}\right)\right|_{x=\pm
L/2}=0.
 \label{ComplexSDBC}
\end{eqnarray}
The initial condition for Eq. (\ref{ComplexSDEq}) is
\begin{equation}
S(x,t=0)=S_0(x).
\label{ComplexSDIC}
\end{equation}

 We can exclude rotations of spin polarization vector (defined by Eq. (\ref{ComplexS})) that are still present in Eq. (\ref{ComplexSDEq}) and Eq. (\ref{ComplexSDBC}) by introducing a complex field $u$ via
\begin{eqnarray}
u(x,t)=e^{-i\eta x}S(x,t) \label{u}.
\end{eqnarray}
Note that the rotation transformation (\ref{u}) is a particular case  of more
general transformations that can be used to remove spin precession~\cite{Tokatly10b}.
 It can be shown~\cite{Slipko11a} that Eq. (\ref{ComplexSDEq}) and Eq. (\ref{ComplexSDBC}) transform into the ordinary heat equation
\begin{equation}
\frac{\partial u}{\partial t}
=D\frac{\partial^2 u}{\partial x^2},
\label{uDEq}
\end{equation}
with boundary condition
\begin{equation}
\left. \left(\frac{\partial u}{\partial x}\pm\frac{1-\gamma}{1+\gamma}\frac{ u}{l}\right)\right|_{x=\pm
L/2}=0,
 \label{uDBC}
\end{equation}
 where $D=l^2/\tau$. Note that in the diffusion
approximation $y$-component of spin polarization, $S_y$, satisfies the same Eq.
(\ref{uDEq}) and Eq. (\ref{uDBC}).

\begin{figure}[t]
 \begin{center}
    \includegraphics[width=7cm]{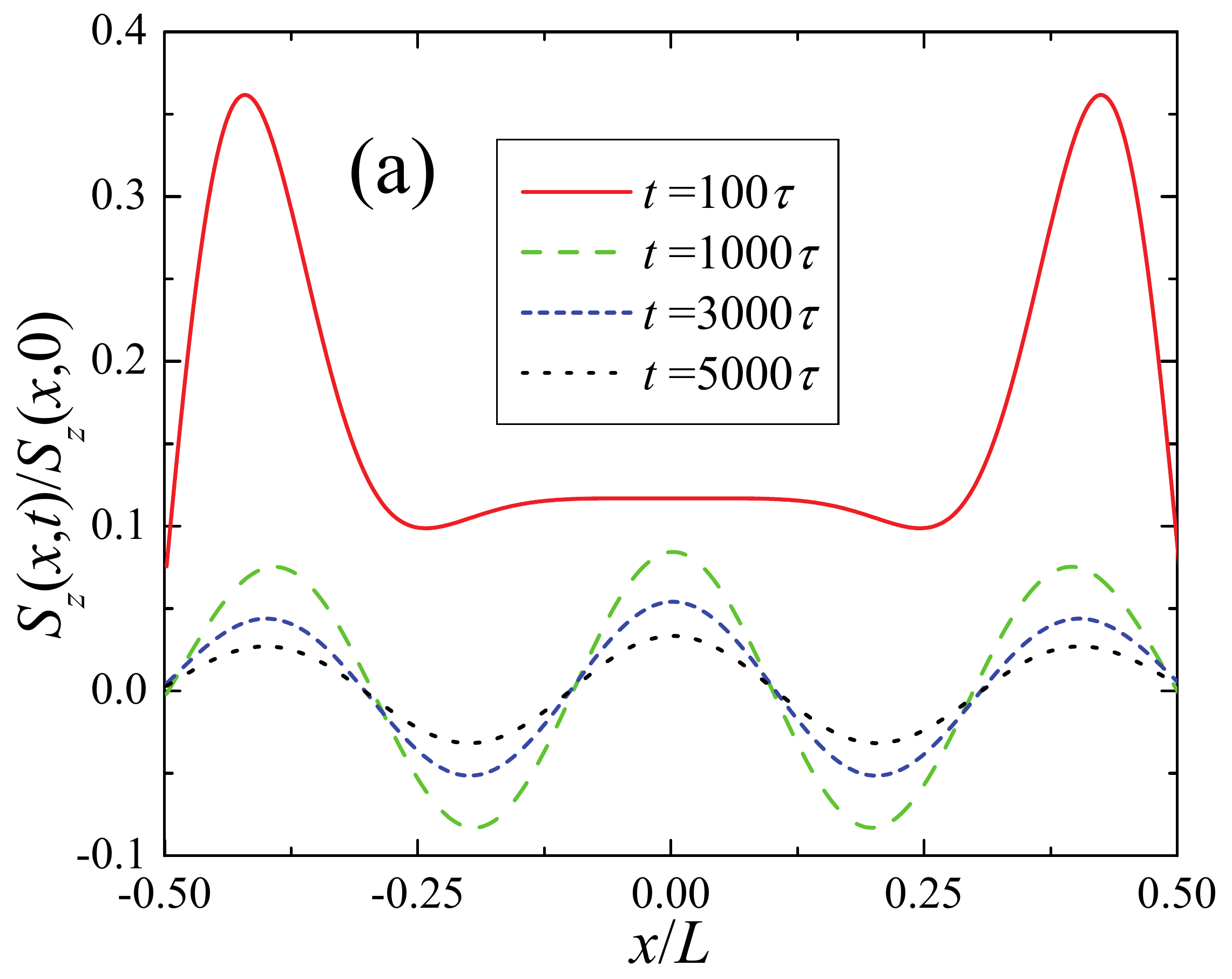}
    \includegraphics[width=7cm]{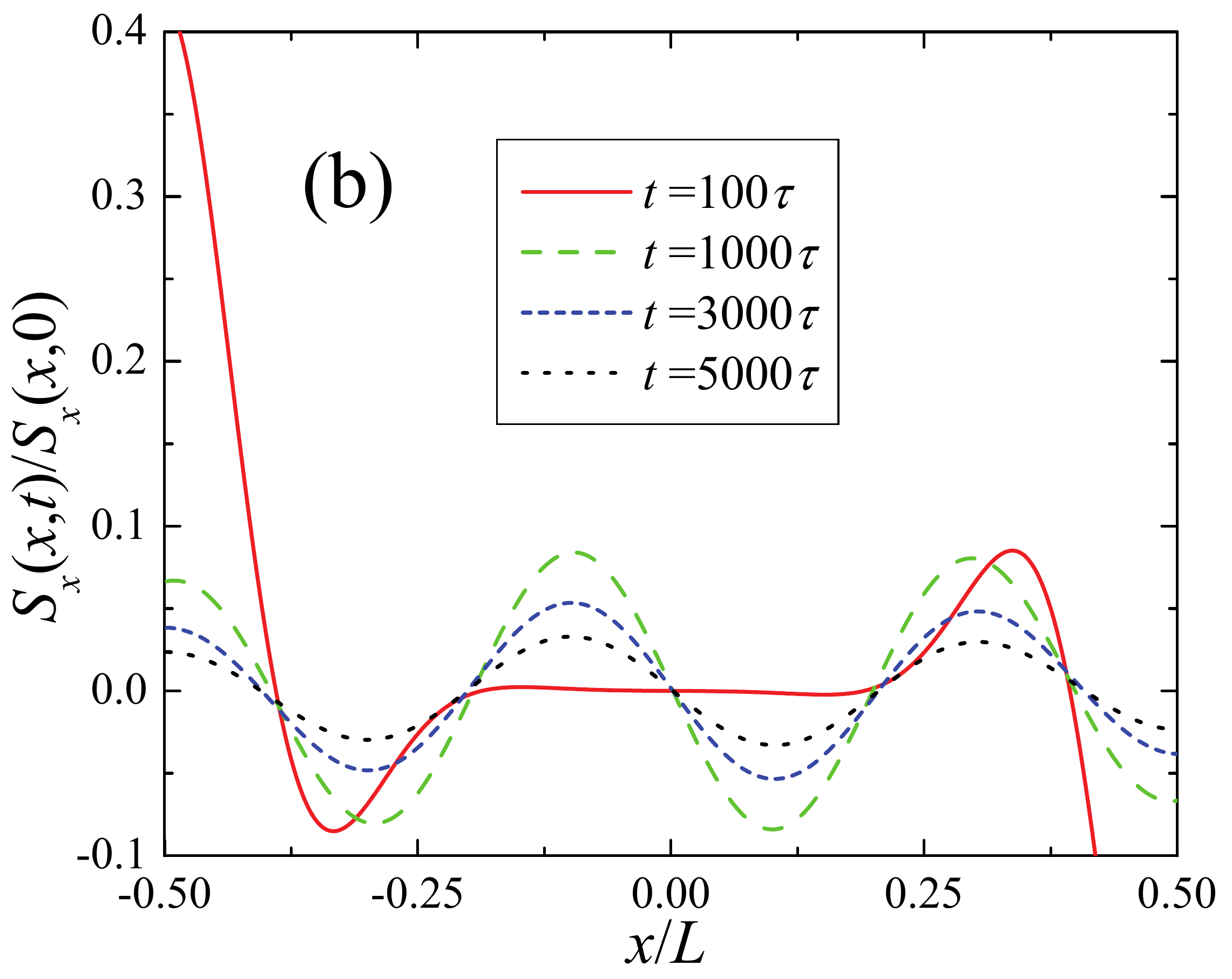}
\caption{\label{fig1} Relaxation of homogeneous spin polarization initially polarized in $z$-direction. (a) and (b) show $z$- and $x$-components of spin polarization density at several moments of time. The wire length $L=100l$, $\eta l=0.1545$, $\gamma=0.97$.
These parameters result in $\kappa=1.31$, and $\xi_1=0.78$. }
\end{center}
\end{figure}


\begin{figure}[t]
 \begin{center}
    \includegraphics[width=7cm]{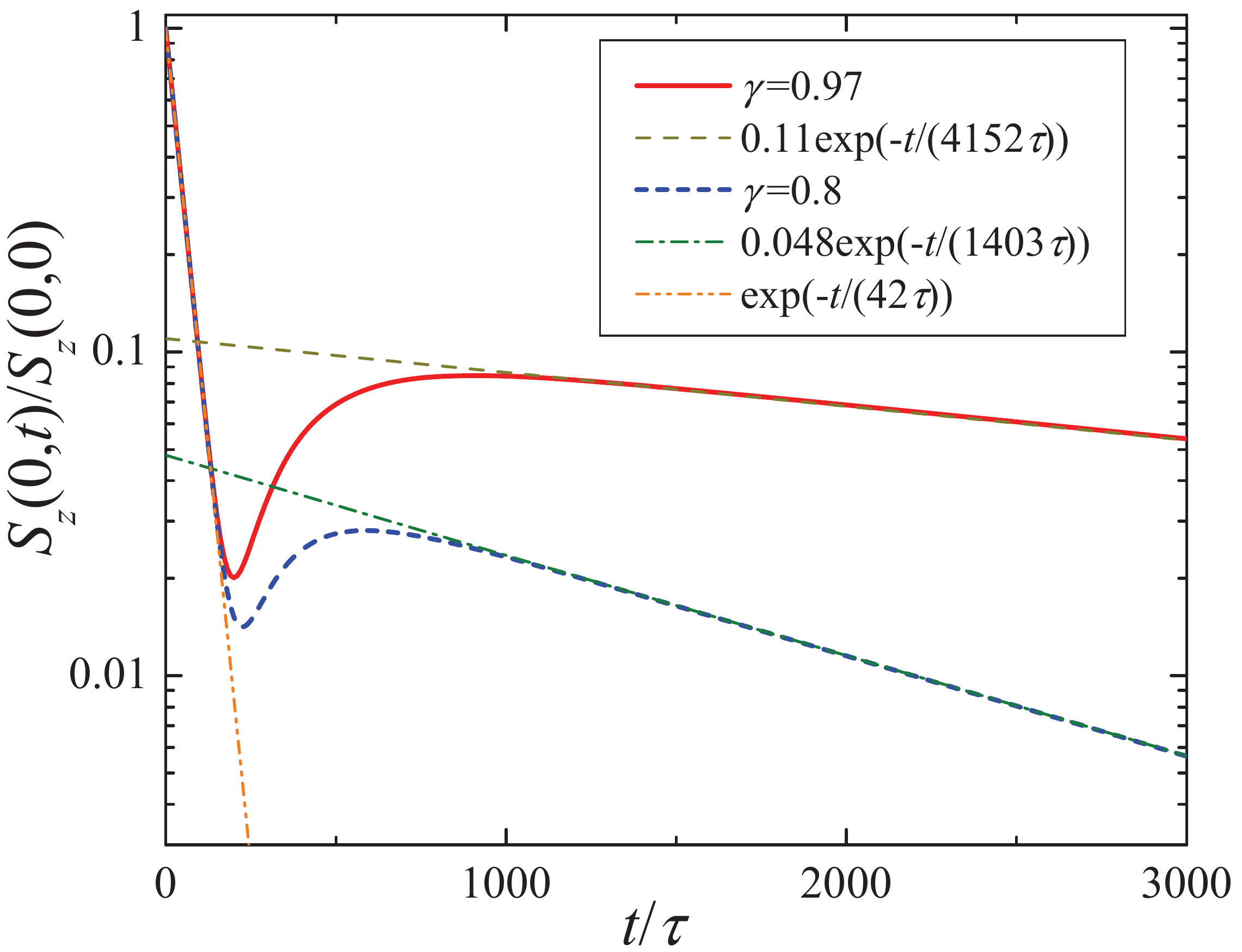}
\caption{\label{fig2} Temporal evolution of the spin polarization density $S_z$ at the wire's center
for the same parameter values as in Fig. \ref{fig1}. The fitting curves are obtained employing numerical values of $\tau_{DP}$ and $\tau_r$ calculated
for these parameter values.
}
\end{center}
\end{figure}

 We find the general solution of Eq. (\ref{uDEq}) with the boundary conditions (\ref{uDBC}) using the standard method of separation of variables. The straightforward application of this method leads to the following expression for the complex spin polarization
\begin{eqnarray}
S&=& \label{SGenSol} \\
 &e^{i\eta x}&
 \sum_{n=1}^{+\infty}
\left(a_ne^{-\frac{4\mu^2_nD}{L^2}t}
\sin
\frac{2 \mu_n x}{L}+b_ne^{-\frac{4\xi^2_nD}{L^2}t}
\cos
\frac{2 \xi_n x}{L}\right),~
\nonumber
\end{eqnarray}
where $\mu_n$ and $\xi_n$ are positive roots of
\begin{eqnarray}
\tan\mu_n=-\kappa\mu_n, \;\;\;\; \textnormal{and}\;\;\;\; \cot\xi_n=\kappa\xi_n,
\label{tanEq}
\end{eqnarray}
and $\kappa=2(1+\gamma)l/[(1-\gamma)L]$.

Thinking about relaxation of an initial spin polarization profile, it is evident that an 'overall' spin relaxation rate is determined by the smallest of the roots of Eqs. (\ref{tanEq}), which is the root $\xi_1$ satisfying the inequality  $0<\xi_1<\pi/2$ (note that $\pi/2<\mu_1<\pi$). We can find explicit expressions for the spin relaxation time $\tau_{r}=L^2 / (4D\xi^2_1)$ in the limiting cases of small and large $\kappa$. Specifically, when $\kappa\ll 1$ (the limit of strong boundary spin relaxation),
\begin{equation}
\xi_1=\frac{\pi}{2}(1-\kappa),\;\;\;\; \textnormal{and} \;\;\;\; \tau_{r}=\frac{L^2\tau}{\pi^2 l^2}(1+2\kappa).
\end{equation}
In the opposite limit, when $\kappa\gg 1$ (the limit of weak boundary spin relaxation), we obtain
\begin{equation}
\xi_1=\frac{6\kappa-1}{6\kappa^{\frac{3}{2}}},\;\;\;\; \textnormal{and} \;\;\;\; \tau_{r}=\frac{(1+\gamma)(1+(3\kappa)^{-1})L\tau}{2(1-\gamma)l}.
\end{equation}

There are two additional time scale in the problem -- the D'yakonov-Perel' spin relaxation time, $\tau_{DP}=\tau / (l\eta)^2 $ and the time of persistent spin helix formation in the absence of boundary spin scattering, $\tau_h$. As these time scales are hidden in the set of relaxation times describing dynamics of the general solution (Eq. (\ref{SGenSol})), we refer to our prior publication~\cite{Slipko11a} where the expression for $\tau_h$ is given, $\tau_h=L^2/(\pi^2 D)$. It is interesting that $\tau_h$ can not exceed $\tau_r$, namely, $\tau_r\geq \tau_h$. Consequently, when $\tau_r \gg \tau_h$, the spin dynamics can be considered as a three-stage process: an initial D'yakonov-Perel' spin relaxation ({\it i}) is followed by spin helix formation ({\it ii}), which is followed by its decay ({\it iii}). Otherwise, when $\tau_r$ and $\tau_h$ are close, the decay stage significantly overlaps with the spin helix formation and thus the stages ({\it ii}) and ({\it iii}) can not be well separated.

In the case of three-stage dynamics, the spin helix amplitude (at the end of the second stage) is defined by the relation between $\tau_{DP}$ and $\tau_h$ times. In particular, if $\tau_h \gg \tau_{DP}$ (or, equivalently, $L\eta\gg\pi$), then the spin helix amplitude is much smaller than the amplitude of the initial spin polarization. In the opposite regime, when $\tau_h \lesssim \tau_{DP}$ (or $L\eta\lesssim\pi$), the spin helix amplitude is comparable to the initial one.

As an application of the above theory, let us consider a specific problem -- the problem of relaxation of initially homogeneous spin polarization pointing in $z$ direction. In this case, the initial condition simply reads $S(x,0)=iS_0$ (see Eq. (\ref{ComplexS})). The coefficients $a_n$ and $b_n$ are obtained by substitution this initial condition into Eq. (\ref{SGenSol}), its multiplication by an appropriate sine or cosine function and subsequent integration over
$x$. This results in
\begin{equation}
\frac{a_n}{S_0}=\frac{2\mu_n\cos\mu_n(\sin (0.5\eta L)+
0.5\eta L\kappa\cos (0.5\eta L))}
{((0.5\eta L)^2-\mu^2_n)(1+\kappa \cos^2\mu_n)},\;\;\;
\label{anHom}
\end{equation}
\begin{equation}
\frac{b_n}{S_0}=i\frac{2\xi_n\sin\xi_n(\cos (0.5\eta L)-
0.5\eta L\kappa\sin (0.5\eta L))}
{(\xi^2_n-(0.5\eta L)^2)(1+\kappa \sin^2\xi_n)}.\;\;\;
\label{bnHom}
\end{equation}

Fig. \ref{fig1} demonstrates dynamics of initially homogeneous spin polarization plotted using Eqs. (\ref{SGenSol}), (\ref{anHom}), (\ref{bnHom}). We emphasize that the initial relaxation at the center is of D'yakonov-Perel' type as, e.g., $S_z$ and $S_x$ in the central region (see $t=100\tau$ curves) are flat.
The spin helix configuration is clearly seen at $t \sim 1000 \tau$. The subsequent slower evolution of spin polarization is caused by the boundary spin scattering. The boundary spin scattering results in smaller amplitudes of spin polarization oscillations closer to the ends and a larger amplitude at the wire's center.

Three stages of spin polarization dynamics, however, are better visualized plotting the magnitude of $S_z$ at the wire's center as a function of time. Fig. \ref{fig2} shows $S_z(0,t)$  for two selected values of boundary spin scattering coefficient $\gamma$. Clearly, both curves contain two intervals of exponential relaxation that are seen on this logarithmic plot as straight lines. The initial interval of  D'yakonov-Perel' spin relaxation ($0<t\lesssim 200\tau$) is not influenced by the boundary spin relaxation. The later, however, determines the character of spin relaxation at long times $750\tau\lesssim t$. The interval of spin helix formation ($200\tau\lesssim t\lesssim 750\tau$)) is located between two regions of exponential evolution mentioned above.


\begin{figure}[tb]
 \begin{center}
    \includegraphics[width=7cm]{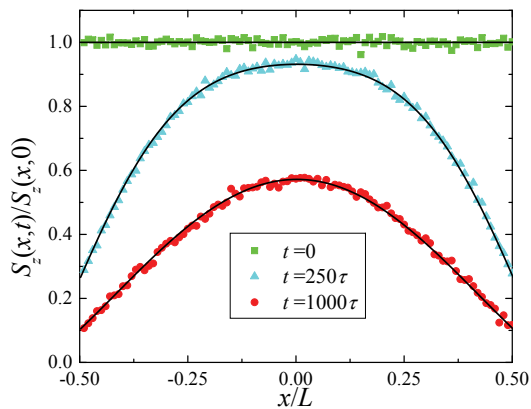}
\caption{\label{fig3} Comparison of $z$-component of spin polarization calculated analytically (straight lines) and numerically using Monte Carlo simulations (dots) for $L=100l$, $\eta l=0.01545$, $\gamma=0.8$.
}
\end{center}
\end{figure}

In order to obtain an additional insight on spin relaxation with boundary spin scattering, we have performed Monte Carlo
simulations using an approach described in Refs. \onlinecite{Kiselev00a} and \onlinecite{Saikin05a}.
This approach is based on a semiclassical description of electron space motion and quantum-mechanical description of spin dynamics.
All main parts of the Monte Carlo simulation algorithm are described in
Refs. \onlinecite{Kiselev00a} and \onlinecite{Saikin05a} and will not be repeated here.
The only novel feature of the code is the boundary spin scattering mechanism implemented in the following way.
When an electron scatters from a boundary, the direction of its spin is inverted with a probability $p=(1-\gamma)/2$. It is not difficult to notice that this part of the algorithm is equivalent to the boundary conditions (Eqs. (\ref{BCmp}), (\ref{BCpm})) for the spin kinetic equation.

There is an excellent agreement between our analytical results and the numerical Monte Carlo simulations. For example, let us consider dynamics of spin relaxation in a finite length wire with boundary spin scattering when the spin precession angle per mean free path is relatively small. We also assume that at the initial moment of time $t=0$, the spin polarization is homogeneous and points in $z$ direction. For this situation, Fig. \ref{fig3} presents a comparison of $S_z$ at different moments of time plotted using Eqs. (\ref{SGenSol}), (\ref{anHom}), (\ref{bnHom}) (smooth curves,) and Monte Carlo simulations (dots). We see that there is an excellent agreement between the analytical and
numerical results.

In conclusion, we have developed a theory of spin relaxation in wires accounting for boundary spin relaxation. For both spin kinetic and diffusion equations appropriate boundary conditions have been derived. Based on this theory, we predict the existence of three (in some cases, however, two) stages of spin dynamics consisting of an initial D'yakonov-Perel' relaxation followed by spin helix formation and its subsequent decay. Experimentally, parameters of boundary spin relaxation can be extracted from both the long-time spin helix decay rate and spin helix shape distortion.

\vspace{1cm}

This work has been partially supported by the University of South Carolina ASPIRE grant 13070-12-29502.

\bibliography{spin}

\end{document}